\documentstyle[11pt,newpasp,twoside,epsfig]{article}
\index{neutron stars!equation of state}

\def\edcomment#1{\iffalse\marginpar{\raggedright\sl#1\/}\else\relax\fi}
\reversemarginpar

\begin{document}
\title{Nuclear pasta structure in hot neutron stars
}
\author{Gentaro Watanabe$^{a,b,c}$, Katsuhiko Sato$^{a,d}$,
Kenji Yasuoka$^{e}$ and Toshikazu Ebisuzaki$^{b}$}
\affil{$^{a}$Department of Physics, University of Tokyo,
Tokyo 113-0033, Japan}
\affil{$^{b}$RIKEN,
Saitama 351-0198, Japan}
\affil{$^{c}$NORDITA, Blegdamsvej 17, DK-2100 Copenhagen \O, Denmark}
\affil{$^{d}$RESCEU,
University of Tokyo, Tokyo 113-0033, Japan}
\affil{$^{e}$Department of Mechanical Engineering, Keio University,
Yokohama 223-8522, Japan}

\begin{abstract}
Structure of cold and hot dense matter at subnuclear densities
is investigated by quantum molecular dynamics (QMD) simulations.
Obtained phase diagrams show that the density of
the phase boundaries between the different nuclear structures
decreases with increasing temperature due to the thermal expansion
of nuclear matter region.
\end{abstract}

At subnuclear densities,
nuclear matter exhibits the coexistence of a liquid phase with a gas phase.
Just below the density where nuclei melt into uniform matter,
it is expected that, at sufficiently low temperatures ($T \ll 1$ MeV),
the energetically favorable configuration
of the mixed phase possesses interesting spatial structures
such as rodlike and slablike nuclei, etc.,
which are called nuclear ``pasta'' (Hashimoto, Seki, \& Yamada 1984;
Ravenhall, Pethick, \& Wilson 1983).
While nuclear ``pasta'' at zero temperature is studied by several authors,
``pasta'' phases at finite temperatures relevant to supernova cores
and crusts of young hot neutron stars
have not been studied yet except for some limited cases
(Lassaut et al. 1987; Watanabe, Iida, \& Sato 2000, 2001, 2003).

In the present work, we study the structure of hot dense matter
at subnuclear densities by QMD.
Simulations of the $(n,p,e)$ system at proton fraction 
$x=0.3$ and 0.5 are performed with 2048 nucleons in a cubic periodic box.
The relativistic degenerate electrons
are regarded as a uniform background
and the Coulomb interaction is calculated by the Ewald method.
The effective Hamiltonian used in this work
is that developed by Maruyama et al. (1998).

\begin{figure}[htbp]
\centerline{\rotatebox{-90}{\psfig{file=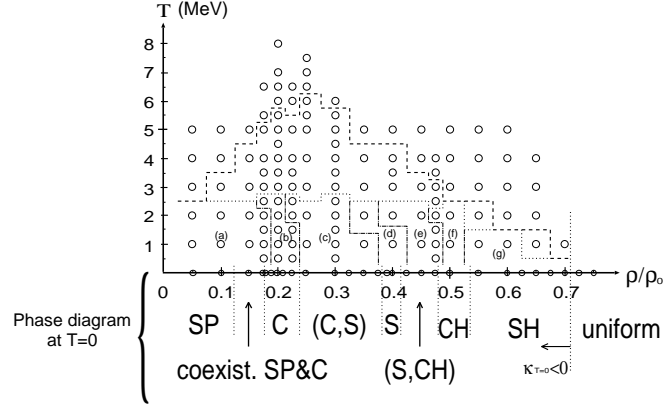,width=5.5cm}}}
\caption{\label{fig1}
  Phase diagram at $x=0.5$.
  The dashed and the dotted lines on the diagram
  show the phase separation line and
  the limit below which the nuclear surface can be identified, respectively.
  The dash-dotted lines are the phase boundaries between
  the different nuclear shapes.
  The symbols SP, C, S, CH and SH stand for
  nuclear shapes,
  i.e., 
  sphere, cylinder, slab, cylindrical hole and spherical hole,
  respectively.
  The parentheses (A,B) show ``spongelike'' intermediate phases
  between A and B-phases
  suggested in our previous works at $T=0$ (Watanabe et al. 2002, 2003).
  They have negative $\chi/V$.
  Simulations have been carried out at points denoted by circles.
  For regions (a)-(g), see text.
}
\end{figure}

We show the resultant phase diagram for $x=0.5$
on the nucleon density $\rho$ vs temperature $T$ plane in Fig.\ \ref{fig1}.
Phase separation line is determined by disappearance of the long-range
correlation of the nucleon distribution detected
by the two-point correlation function.
Nuclear surface is identified by plateau of
the Euler characteristic density $\chi/V$ as a function of threshold density
for isodensity surface.
In the density region of interest, nuclear surface cannot be observed
typically at $T \ga 3$ MeV, even in the phase-separating region.
At temperatures where nuclear surface can be identified,
we characterise the nuclear shape changes with increasing the density
by using averaged mean curvature $\langle H \rangle$ and $\chi/V$:
(a) $\langle H \rangle > 0,\ \chi/V > 0$ $\rightarrow$
(b) $\langle H \rangle > 0,\ \chi/V = 0$ $\rightarrow$
(c) $\langle H \rangle > 0,\ \chi/V < 0$ $\rightarrow$
(d) $\langle H \rangle = 0,\ \chi/V = 0$ $\rightarrow$
(e) $\langle H \rangle < 0,\ \chi/V < 0$ $\rightarrow$
(f) $\langle H \rangle < 0,\ \chi/V = 0$ $\rightarrow$
(g) $\langle H \rangle < 0,\ \chi/V > 0$ $\rightarrow$
uniform.
Critical point for the phase separation locates at $\rho \sim 0.25\rho_0$
and $T \ga 6$ MeV.
It is noted that the density of each phase boundary
between the different structures decreases as $T$ increases,
which is due to the thermal expansion of the nuclear matter region.

Our result suggests that the ``pasta'' phases, and in addition,
``spongelike'' phases with negative values of $\chi/V$
exist in hot neutron star crusts and supernova cores
at $T\la 3$ MeV.

\begin{references}

\reference Hashimoto, M., Seki, H., \& Yamada, M.
  1984, Prog.\ Theor.\ Phys., 71, 320
\reference Lassaut, M., Flocard, H., Bonche, P., Heenen, P. H., \& Suraud, E.
  1987, \aap, 183, L3
\reference Maruyama, T., Niita, K., Oyamatsu, K., Maruyama, T.,
  Chiba, S., \& Iwamoto, A.
  1998, \prc, 57, 655
\reference Ravenhall, D. G., Pethick, C. J., \& Wilson, J. R.
  1983, \prl, 50, 2066
\reference Watanabe, G., Iida, K., Sato, K.
  2000, Nucl.\ Phys. A676, 455;
  2001, Nucl.\ Phys. A687, 512; 2003, Nucl.\ Phys. A726, 357
\reference Watanabe, G., Sato, K., Yasuoka, K., \& Ebisuzaki, T.
  2002, \prc, 66, 012801(R);
  2003, \prc, 68, 035806 (nucl-th/0308007)
%
\end{references}
\end{document}